\documentclass{article}
\usepackage{codefuse_tech_report}
\usepackage[colorlinks = true,
            linkcolor = blue,
            urlcolor  = blue,
            citecolor = blue,
            anchorcolor = blue]{hyperref}   
\usepackage{microtype}
\usepackage{xurl}
\usepackage{booktabs}
\usepackage{enumitem}
\usepackage{multicol}
\usepackage{CJKutf8}
\usepackage{siunitx}
\usepackage{floatflt}
\usepackage{graphicx}
\usepackage{wrapfig}
\usepackage{authblk}
\usepackage{lipsum}
\usepackage{algorithm}
\usepackage{algorithmicx}
\usepackage{algpseudocode}
\usepackage{subcaption}
\usepackage{multirow}
\usepackage{pifont}  

\algnewcommand{\LeftComment}[1]{\Statex \(\triangleright\) #1}

\usepackage{array}
\usepackage{amsmath}
\usepackage{amssymb}
\usepackage{mathtools}
\usepackage{amsthm}

\usepackage[capitalize,noabbrev]{cleveref}
\usepackage{adjustbox} 

\theoremstyle{plain}

\theoremstyle{definition}

\theoremstyle{remark}

\colmfinalcopy

\usepackage[utf8]{inputenc}
\usepackage[T1]{fontenc}
\usepackage{caption} 
\usepackage{adjustbox} 
\usepackage{fontawesome5}

\usepackage{tcolorbox}
\tcbuselibrary{skins,breakable}

\tcbuselibrary{skins}

\usepackage{color}
\usepackage{xcolor}
\usepackage{soul} 

\definecolor{tred}{RGB}{251, 130, 132}
\definecolor{torange}{RGB}{247, 162, 116}
\definecolor{tyellow}{RGB}{251, 218, 140}
\definecolor{tgreen}{RGB}{127, 204, 181}
\definecolor{tblue}{RGB}{89, 177, 215}
\definecolor{insightblue}{RGB}{162, 210, 255}
\definecolor{questionred}{RGB}{255, 175, 204}

\usepackage{multirow}
\usepackage{graphicx}
\usepackage{threeparttable}
\newcommand{\tabincell}[2]{\begin{tabular}{@{}#1@{}}#2\end{tabular}}
\newcommand{\system}{\textit{\sloppy{CodeFuse-CR-Bench}\@}}
\graphicspath{{./figure}}
\usepackage{tikz}
\usepackage{wasysym}

\title{CodeFuse-CR-Bench: A Comprehensiveness-aware Benchmark for End-to-End Code Review Evaluation in Python Projects}

\author{%
Hanyang Guo\thanks{Equal Contribution.}$^{\phantom{*},1}$
~~Xunjin Zheng$^{*,1}$
~~Zihan Liao$^{1}$
\\

\vspace{-6pt}
\bf
~~Hang Yu$^{1,\dagger}$
~~Peng DI$^{1,2,}$\thanks{Correspondence to: Hang Yu \textless hyu.hugo@antgroup.com\textgreater ~and Peng Di \textless dipeng.dp@antgroup.com\textgreater.} 
~~Ziyin Zhang$^{1}$
~~Hong-Ning Dai$^{3}$

\vspace{10pt}
$^1$Ant Group\ \ \ $^2$UNSW Sydney\ \ \ $^3$Hong Kong Baptist University\\
\vspace{10pt}
\hspace{-10pt}\faGithub ~\url{https://github.com/codefuse-ai/SWE-CARE}\\
\hspace{-10pt}~~~~~~~~\includegraphics[width=1em,height=1em]{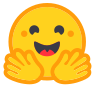} ~\url{https://huggingface.co/datasets/inclusionAI/SWE-CARE}\\
}

\colmfinalcopy 

\begin{document}

\maketitle


\begin{abstract}
Automated code review (CR) is a key application for Large Language Models (LLMs), but progress is hampered by a ``reality gap'': existing benchmarks evaluate models on isolated sub-tasks using simplified, context-poor data. This fails to reflect the holistic context-rich nature of real-world CR. To bridge this gap, we introduce \system{}, the first \textbf{comprehensiveness-aware} benchmark for repository-level CR evaluation. \system{} comprises 601 high-quality instances from 70 Python projects covering nine Pull-Request (PR) problem domains, where each instance provides rich, multi-faceted context including the associated issue, PR details, and repository state, enabling end-to-end evaluation. Beyond superficial metrics, we also propose a novel evaluation framework that combines rule-based checks for location and syntax with model-based judgments of review quality. We present the first large-scale assessment of state-of-the-art LLMs on this comprehensive CR task. Our results establish crucial baselines and reveal that (1) no single LLM dominates all aspects of CR; (2) Gemini 2.5 Pro achieves the highest comprehensive performance; and (3) different LLMs exhibit varying robustness to redundant context. These findings highlight the necessity of holistic, multi-dimensional evaluation and provide actionable insights for advancing truly intelligent yet practical CR assistants.
\end{abstract}

\section{Introduction}

Code Review (CR) is a core practice in modern software development that aims to improve code quality and identify defects through collaborative inspection~\citep{Alberto2013}. As Large Language Models (LLMs) increasingly automate complex software engineering tasks, their application to CR holds immense promise for improving software quality and developer productivity. However, the development of such sophisticated tools is fundamentally constrained by how we measure their performance. Current benchmarks, while valuable, evaluate models in isolated, decontextualized settings, creating a significant and growing ``reality gap'' between measured performance and real-world efficacy.

The core of this problem lies in a failure to capture the \textbf{comprehensiveness} of the CR process. Real-world CR is not a simple text-matching exercise; instead it is a holistic reasoning task that requires a deep understanding of context. This disconnection manifests in three critical limitations of existing automated CR research, as demonstrated in Table~\ref{tab:benchmark difference}:
\begin{enumerate}[leftmargin=*,labelindent=0pt,noitemsep,topsep=0pt]
    \item \textbf{Task Fragmentation}: The cognitive process of a human reviewer—understanding the initial problem, locating potential issues in a code change, and formulating a coherent review—is often broken down into isolated sub-tasks like comment generation or code refinement. This fragmentation prevents the evaluation of end-to-end reasoning, a crucial capability for a truly useful automated reviewer.
    \item \textbf{Context Poverty}: Existing benchmarks typically provide only small, self-contained code snippets and strip away the rich context that is essential for meaningful review, such as the Pull Request (PR) description, the linked issue report, and the broader repository structure. Without this context, a model cannot grasp the intent behind a change, making its review superficial.
    \item \textbf{Evaluation Narrowness}: Evaluation metrics are often inherited from natural language processing (NLP) tasks (e.g., BLEU). These metrics reward superficial textual similarity but fail to assess the substantive quality of a review. They cannot distinguish a technically insightful suggestion from a syntactically similar but incorrect one, nor can they verify if a review comment is even placed at the correct location in the code.
\end{enumerate}

\begin{table}
\caption{A comparative analysis of \system{} and other prominent code review (CR) and issue-solving benchmarks, illustrating the ``comprehensiveness gap'' in existing CR evaluation. The table demonstrates the ``task fragmentation'' and ``context poverty'' of prior benchmarks, which often omit crucial information necessary for a holistic review. \system{} addresses this gap by providing rich, repository-level context, encompassing PR and issue details, commit history, and the complete patch for review. Legend: \CIRCLE~= Present, \Circle~ = Absent, \LEFTcircle~= Partially Present. Refer to Section~\ref{sec:benchmark characteristics} for more details.}
\label{tab:benchmark difference}
\resizebox{\linewidth}{!}{
\begin{threeparttable} 
  \begin{tabular}{cccccccccc}
    \toprule
    \textbf{Benckmarks}&
    \textbf{\system{}}&
    {\tabincell{c}{\textbf{Trans-Review}\\\citep{Tufano2021}}}&{\tabincell{c}{\textbf{AutoTransform}\\\citep{Patanamon2022}}}&
    {\tabincell{c}{\textbf{T5-Review}\\\citep{Tufano2022}}}&
    {\tabincell{c}{\textbf{CodeReviewer}\\\citep{Li2022}}}&
    {\tabincell{c}{\textbf{AUGER}\\\citep{Li2022_1}}}&
    {\tabincell{c}{\textbf{SWE-Bench}\\\citep{jimenez2024}}}&
    {\tabincell{c}{\textbf{Multi-SWE-Bench}\\\citep{zan2025}}}&
    {\tabincell{c}{\textbf{FAUN-Eval}\\\citep{hu2024}}}\\
    \midrule
    \textbf{Instance ID}& \CIRCLE & \Circle & \Circle & \Circle & \Circle & \Circle & \CIRCLE & \CIRCLE & \Circle\\
    \textbf{Owner/Repo}& \CIRCLE & \Circle & \Circle & \Circle & \Circle & \Circle & \CIRCLE & \CIRCLE & \CIRCLE\\
    \textbf{Language}& \CIRCLE & \Circle & \Circle & \Circle & \Circle & \Circle & \Circle & \CIRCLE & \CIRCLE\\
    \textbf{Pull No.}& \CIRCLE & \Circle & \Circle & \Circle & \Circle & \Circle & \CIRCLE & \CIRCLE & \CIRCLE\\
    \textbf{Title}& \CIRCLE & \Circle & \Circle & \Circle & \Circle & \Circle & \Circle & \Circle & \CIRCLE\\
    \textbf{Created at}& \CIRCLE & \Circle & \Circle & \Circle & \Circle & \Circle & \CIRCLE & \CIRCLE & \CIRCLE\\
    \textbf{Base Commit}& \CIRCLE & \Circle & \Circle & \Circle & \Circle & \Circle & \CIRCLE & \CIRCLE & \CIRCLE\\
    \textbf{Body}& \CIRCLE & \Circle & \Circle & \Circle & \Circle & \Circle & \Circle & \Circle & \CIRCLE\\
    \textbf{ISP}\tnote{1}& \CIRCLE & \Circle & \Circle & \Circle & \Circle & \Circle & \CIRCLE & \CIRCLE & \Circle\\
    \textbf{Hint Text}& \CIRCLE & \Circle & \Circle & \Circle & \Circle & \Circle & \CIRCLE & \CIRCLE & \Circle\\
    \textbf{Resolved Issue No.}& \CIRCLE & \Circle & \Circle & \Circle & \Circle & \Circle & \CIRCLE & \CIRCLE & \Circle\\
    \textbf{CPR}\tnote{2}& \CIRCLE & \LEFTcircle: MLP\tnote{6} & \LEFTcircle:MLP & \LEFTcircle: MLP & \LEFTcircle: DHLP & \LEFTcircle: MLP & \CIRCLE & \CIRCLE & \Circle\\
    \textbf{Head Commit}& \CIRCLE & \Circle & \Circle & \Circle & \Circle & \Circle & \Circle & \Circle & \Circle\\
    \textbf{HCM}\tnote{3}& \CIRCLE & \Circle & \Circle & \Circle & \Circle & \Circle & \Circle & \Circle & \Circle\\
    \textbf{Problem Domain}& \CIRCLE & \Circle & \Circle & \Circle & \Circle & \Circle & \Circle & \Circle & \Circle\\
    \textbf{Difficulty}& \CIRCLE & \Circle & \Circle & \Circle & \Circle & \Circle & \Circle & \Circle & \Circle\\
    \textbf{RCT}\tnote{4}& \CIRCLE & \CIRCLE & \Circle & \CIRCLE & \CIRCLE & \CIRCLE & \Circle & \Circle & \CIRCLE\\
    \textbf{Diff Hunk}& \CIRCLE & \Circle & \Circle & \Circle & \CIRCLE & \Circle & \CIRCLE & \CIRCLE & \Circle\\
    \textbf{Review Effort}& \CIRCLE & \Circle & \Circle & \Circle & \Circle & \Circle & \Circle & \Circle & \Circle\\
    \textbf{MCP}\tnote{5}& \CIRCLE & \Circle & \Circle & \Circle & \Circle & \Circle & \CIRCLE & \CIRCLE & \CIRCLE\\
    \textbf{Merge Commit}& \CIRCLE & \Circle & \Circle & \Circle & \Circle & \Circle & \CIRCLE & \CIRCLE & \Circle\\ 
  \bottomrule
\end{tabular}
\begin{tablenotes}[para]
\item[1] Issue Problem Statement \item[2] Commit Patch to Review
\item[3] Head Commit Message
\item[4] Review Comment Text
\item[5] Merge Commit Patch
\item[6] Method-level Patch
\item[7] Diff-Hunk-level Patch

\end{tablenotes}
\end{threeparttable}
}
\end{table}

\begin{wrapfigure}{l}{0.5\textwidth}
\centering
  \includegraphics[width=\linewidth]{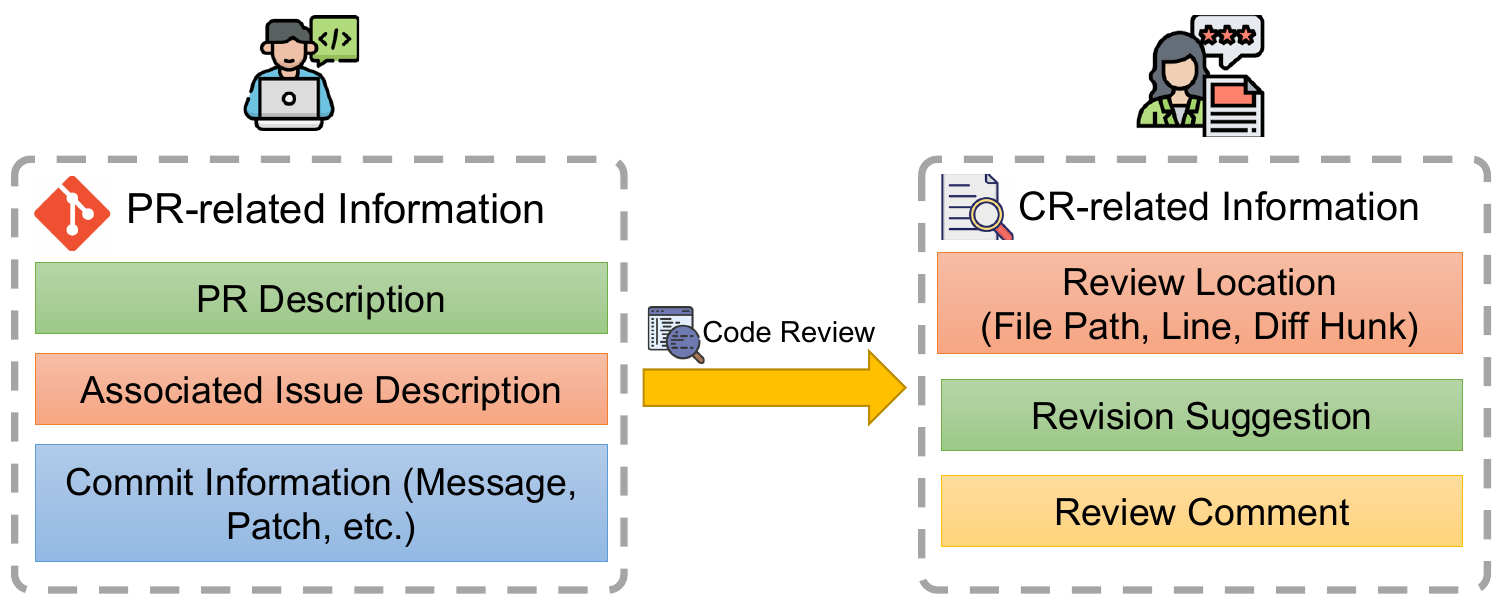}
  \caption{CR Process}
  \label{CR_process}
\end{wrapfigure}

In contrast, as shown in Figure~\ref{CR_process}, a typical CR process is a holistic reasoning task. The code reviewer receives a CR request and the corresponding PR information, which includes PR-related information such as the PR description and associated issue data. Based on this rich context, the reviewer performs the review by writing comments and revision suggestions, which constitute the CR-related information. 

To bridge the gap between this reality and current evaluation methods, we introduce \textbf{\system{}}, a \textbf{C}omprehensiveness-\textbf{A}ware benchmark for \textbf{R}epository-level \textbf{E}valuation of code review. \system{} is designed from the ground up to model this full CR workflow. It comprises 601 high-quality instances curated from 70 real-world Python open-source projects. Each instance is a rich, self-contained snapshot of a real review task, encompassing basic information, PR-related information, CR-related information, and repository-level context. This multi-faceted context enables models to engage in the kind of holistic reasoning that developers perform daily.

Furthermore, to move beyond narrow, syntax-focused metrics, we designed a comprehensive evaluation framework for \system{}. This framework integrates both fine-grained \textbf{rule-based metrics} (to assess location accuracy and semantic similarity) and holistic \textbf{model-based evaluation} (which uses the reward model and advanced LLMs-as-judges to score the overall quality, relevance, and correctness of a review). Using this benchmark and framework, we conduct the first large-scale assessment of state-of-the-art LLMs on the comprehensive CR task. Our results establish crucial baselines and reveal the current capabilities and limitations of LLMs when faced with the complexities of real-world code review.

In summary, this paper makes the following contributions:
\begin{itemize}[leftmargin=*,labelindent=0pt,noitemsep,topsep=0pt]
    \item We identify and characterize the "comprehensiveness gap" in current CR research, highlighting how task fragmentation, context poverty, and narrow evaluation metrics hinder progress.
    \item We introduce \textbf{\system{}}, the first comprehensiveness-aware CR benchmark that provides rich, repository-level context to enable the evaluation of end-to-end CR tasks across nine distinct problem domains.
    \item We propose a novel, multi-faceted evaluation framework that combines rule-based precision with model-based quality assessment to provide a more holistic measure of CR performance.
    \item We conduct an extensive empirical study on multiple state-of-the-art LLMs, providing the first robust baseline for comprehensive CR and offering insights into future research directions.
\end{itemize}

\section{Related Work}

\subsection{CR-related Tasks}\label{sec:CR related work}
Previously, the vast majority of benchmarks and approaches were constructed to improve CR performance~\citep{jiang2025}. Tufano et al.~\citep{Tufano2021} proposed Trans-Review, which adopted deep learning models to partially automate specific CR tasks. They trained two models to implement two CR sub-tasks: (i) code revision before review and (ii) code revision after review. Based on this prior work, they~\citep{Tufano2022} updated the models from deep learning models to the pre-training model T5 and named it T5-Review. The results demonstrated that T5-Review can outperform previous deep learning models for automating CR tasks. Thongtanunam et al.~\citep{Patanamon2022} proposed AutoTransform, which leverages a Byte-Pair Encoding (BPE) approach to handle new tokens and a Transformer-based Neural Machine Translation architecture to handle long sequences. It can be used in the task of code revision before review. Zhou et al.~\citep{zhou2023} evaluated the above three automatic CR tools and pre-trained models for three processes in CR: code revision before review, review comment generation, and code revision after review. The results show that a general-purpose pre-trained model CodeT5 can outperform other models in most cases. Li et al.~\citep{Li2022_1} proposed a review comments generator with pre-training models, which is called AUGER. They collected empirical review data from 11 notable Java projects and constructed a dataset of 10,882 code changes to evaluate the performance of the proposed approach. Li et al.~\citep{Li2022} proposed a pre-trained model that utilized four pre-training tasks tailored specifically for the CR scenario, named CodeReviewer. They focused on three key tasks related to CR activities, including code change quality estimation, review comment generation, and code refinement to evaluate the model. They also constructed a high-quality benchmark dataset based on our collected data for these three tasks and conducted comprehensive experiments on it. The experiments demonstrated the SOTA results.
Additionally, some LLM-based CR approaches were proposed. Guo et al.~\citep{Guo2024} conducted the first empirical study to explore the capabilities of ChatGPT in CR tasks, specifically focusing on automated code revision after reviews. They constructed a new CR dataset with high quality based on the existing benchmark CodeReview~\citep{Li2022}. A SOTA CR tool~\citep{Li2022} was selected as a baseline. The research study provided insights into the potential of ChatGPT in automating the CR process.

As an important upstream sub-task of CR, issue-solving has also been extensively studied. The content of issue-solving can provide more complete contexts for CR. Some issue-solving benchmarks had been proposed. Jimenez et al.~\citep{jimenez2024} proposed a benchmark, SWE-Bench, that evaluates
LLMs in resolving an issue (typically a bug report or a feature request) submitted to popular Python GitHub repositories. Zan et al.~\citep{zan2025} enlarged the SWE-Bench dataset by adding other programming languages' issue-solving and named it Multi-SWE-bench. Hu et al.~\citep{hu2024} proposed FAUN-Eval, a benchmark specifically designed to evaluate the fine-grained issue-solving capabilities of LLMs. It can be used to systematically assess LLMs across three distinct tasks: Question-Answer (QA), fault localization, and code editing. 

However, either the above approaches or the benchmarks focus on the sub-tasks of CR. The design and construction are comprehensiveness-unaware. To fill this gap, we construct a comprehensiveness-aware benchmark and evaluation metrics for CR. We conduct an empirical study on some SOTA LLMs based on this benchmark to evaluate their performance in comprehensive CR tasks.

\subsection{LLM for Software Engineering Tasks}
Recently, LLMs have demonstrated revolutionary performance improvements in almost all software-engineering-related tasks. Regarding general LLMs, the GPT series~\citep{Liang2024}, Claude, Gemini, and others had demonstrated powerful code generation, summarization, and program repair through training on large corpora containing code~\citep{Feng2024, Cao2024,Wang2025AreI,zhao2023rightpromptsjobrepair,fan2023staticcodeanalysisai}. Specifically, a systematic comparative analysis was conducted on three advanced LLMs, including ChatGPT (O1), DeepSeek (R1), and Gemini (2.0 Flash thinking), for Python code generation, evaluating their performance in correctness, code quality, and computational efficiency. Each of the three LLMs has its own strengths and limitations. Their findings underscored the inherent trade-offs between efficiency, accuracy, and quality in AI-generated code. Sobo et al.~\citep{Sobo2025} investigated the effectiveness of LLMs in generating code for Human-Robot Interaction applications. They compared the performance among ChatGPT 3.5, Gemini 1.5 Pro, and Claude 3.5 Sonnet. The study highlighted the rapid advancement in LLM capabilities for specialized programming tasks while also identifying persistent challenges in spatial reasoning and adherence to specific constraints. Zhang et al.~\citep{Zhang2024} evaluated the capability of advanced LLMs, including ChatGPT-4 and Claude, in fixing memory corruption vulnerabilities in real-world C/C++ code. They analyzed both the strengths and limitations of LLMs in automated program repair on genuine code. Sun et al.~\citep{Sun2025} conducted the examination of prevalent automated evaluation methods for assessing the quality of summaries generated by LLMs and found that the results of the GPT-4 evaluation method are most closely aligned with human evaluation. They also discussed the limitations of LLMs in generating summarization in logic programming languages. All the software-engineering-related tasks mentioned above are highly relevant to CR in technical terms. Therefore, it is possible for LLMs to perform comprehensive CR tasks. In our paper, we select several representative LLMs and evaluate their comprehensive CR capabilities using our constructed benchmark and designed evaluation metrics.

\section{\system{} Benchmark}
Having established the "comprehensiveness gap" in current CR research, this section details the design and construction of \system{}.
We present the benchmark overview, benchmark construction pipeline, and benchmark characteristics in Section~\ref{sec:overview}, Section~\ref{sec:benchmark construction}, and Section~\ref{sec:benchmark characteristics}, respectively.

\begin{figure}[h]
  \centering
  \includegraphics[width=0.9\linewidth]{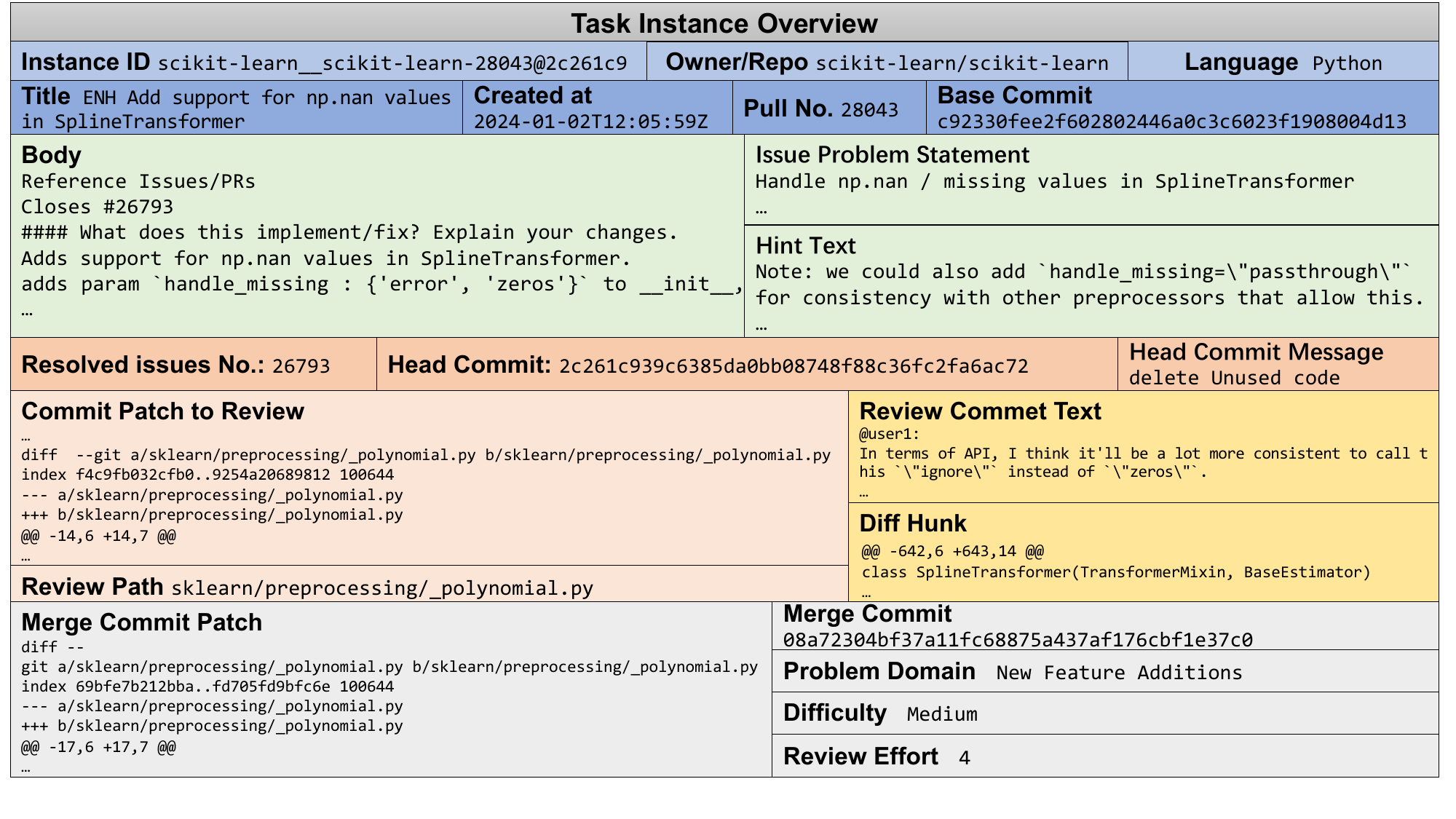}
  \caption{The Overview of A Typical CR Task Instance}
  \label{fig:task instance}
\end{figure}

\subsection{Benchmark Overview}\label{sec:overview}
\system{} comprises 601 Python CR task instances, each carefully curated to reflect real-world development scenarios. Fig.~\ref{fig:task instance} illustrates an overview of a typical CR task instance. Each instance is associated with 22 structured fields, as summarized in Table~\ref{tab:details of fields}.
These instances can be systematically categorized into the following four types to support comprehensive CR: 
\begin{itemize}[leftmargin=*,labelindent=0pt,noitemsep,topsep=0pt]
\item \textbf{Basic information}: \texttt{\small Instance ID}, \texttt{\small Owner/Repo}, and \texttt{\small Language}, providing fundamental identification of the task.
\item \textbf{PR-related information}: Encompasses metadata critical to understanding the change intent and context, including \texttt{\small Pull No.}, \texttt{\small Title}, \texttt{\small Created at}, \texttt{\small Base Commit}, \texttt{\small Body}, \texttt{\small Issue Problem Statement}, \texttt{\small Hint Text}, \texttt{\small Resolved issue No.}, \texttt{\small Commit Patch to Review}, \texttt{\small Head Commit}, \texttt{\small Head Commit Message}, \texttt{\small Problem Domain}, and \texttt{\small Difficulty}.
\item \textbf{CR-related information}: Captures the review process itself, including \texttt{\small Review Comment Text}, \texttt{\small Diff Hunk}, \texttt{\small Review Path}, and \texttt{\small Review Effort}, enabling analysis of reviewer behavior and feedback quality.
\item \textbf{Repository-Level Context Information}: Provides broader project-level context necessary for cross-file reasoning and impact analysis, including \texttt{\small Merge Commit Patch} and \texttt{\small Merge Commit}.
\end{itemize}

\begin{table}
  \caption{Details of Fields in CR Task Instance}
  \label{tab:details of fields}
  \resizebox{\linewidth}{!}{
  \begin{tabular}{llp{9.5cm}}
    \toprule
    \textbf{Category}&\textbf{Field}&\textbf{Description}\\
    \midrule
    \multirow{3}{*}{Basic Information} & \texttt{\small Instance ID} & A formatted instance identifier, which is named as \texttt {\small repo\_owner\_\_repo\_name-PR-number@commit\_hash\_prefix}.\\
    &\texttt{\small Owner/Repo} & Repository owner and repository name.\\
    &\texttt{\small Language} & Repository programming language.\\
    \midrule
    \multirow{13}{*}{PR-related Information} & \texttt{\small Pull No.} & PR No. that this task instance is from.\\
    & \texttt{\small Title} & PR title.\\
    & \texttt{\small Created at} & PR creation date.\\
    & \texttt{\small Base Commit} & Commit hash in this repository before the PR is applied.\\
    & \texttt{\small Body} & PR description body.\\
    & \texttt{\small Issue Problem Statement} & Issue(s) title and body.\\
    & \texttt{\small Hint Text} & Comments made on the issue(s) before the creation of the commit to review of the solution PR.\\
    & \texttt{\small Resolved issue No.} & The number of the issue(s) solved by the PR.\\
    & \texttt{\small Commit Patch to Review} & The target commit patch in the PR to review.\\
    & \texttt{\small Head Commit} & The commit hash of the target commit patch in the PR.\\
    & \texttt{\small Head Commit Message} & The commit message of the target commit patch.\\
    
    & \texttt{\small Problem Domain} & The nine problem domains where the issue problem statement belongs (Section~\ref{sec:feature labeling}).\\
    & \texttt{\small Difficulty} & PR task implementation difficulty.\\
    \midrule
    \multirow{4}{*}{CR-related Information} & \texttt{\small Review Comment Text} & The text of the review comment.\\
    & \texttt{\small Diff Hunk} & The diff hunk where the review comment is located\\
    & \texttt{\small Review Path} & The file path where the review comment is located.\\
    & \texttt{\small Review Effort} & The review effort for a CR task.\\
    \midrule
    \multirow{2}{*}{{\tabincell{l}{Repository-Level Context\\Information}}} & \texttt{\small Merge Commit Patch} & The gold patch generated by the PR to resolve the issue.\\
    & \texttt{\small Merge Commit} & The merged commit hash of the PR.\\
  \bottomrule
\end{tabular}
}
\end{table}

\subsection{Benchmark Construction Pipeline}\label{sec:benchmark construction}
\begin{figure}[h]
  \centering
  \includegraphics[width=\linewidth]{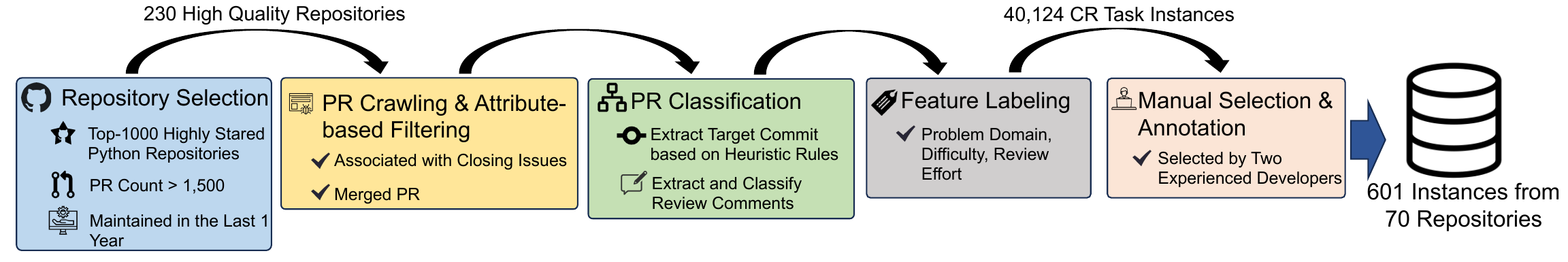}
  \caption{\system{} Construction Pipeline}
  \label{fig:pipeline}
\end{figure}

As depicted in Fig.~\ref{fig:pipeline}, the construction pipeline of \system{} consists of five steps, namely, (1) Repository Selection; (2) PR Crawling and Attribute-based Filtering; (3) PR Classification; (4) Feature Labeling; and (5) Manual Selection \& Annotation. We next briefly elaborate on them. 

\subsubsection{Repository Selection}
To make \system{} more representative, we adopt a strict approach to selecting CR task instances from various open-source repositories. We focus on Python as it is one of the most popular languages on GitHub, possessing a mature and diverse open-source ecosystem. This ensures a rich source of projects with standardized and high-quality CR practices, which is essential for our benchmark's validity. Specifically, we first search for the top 1,000 starred Python repositories on GitHub\footnote{\url{https://github.com/}}, a Git-based code hosting and collaboration platform. A higher star rating indicates that the repository has better popularity. To ensure the quality of the repository and acquire a large pool of potential CR data, we further sort the projects by the number of PRs and filter out projects with less than 1,500 PRs, referring to Li's study~\citep{Li2022}. Additionally, we only keep the repositories that were maintained in the last year (from 2024-08-15 to 2025-08-15), i.e., the repositories that had commit or PR records over the past year. The process mentioned above aims to keep only active projects and remove repositories that are forked from other repositories, as the PR number is not inherited. This process yielded 230 Python projects that meet our criteria for activity and maturity. 

\subsubsection{PR Crawling \& Attribute-based Filtering}
In this step, we aim to crawl high-quality PR data. To ensure the collected PR data contains high-quality CR, we filter it based on the following attributes: (1) We only include PRs that have at least 1 closing issue reference; (2) We only collect the PR that is merged into the main branch. A ``Merged'' status indicates that the code changes associated with the PR were accepted and incorporated into its parent repository. We conduct this process by using GitHub GraphQL API\footnote{\url{https://docs.github.com/en/graphql}}. The PRs selected based on these attributes undergo rigorous CR, thereby ensuring that the CR data selected under these PRs is of high quality~\citep{Zheng2025}.

\begin{table}
  \caption{Heuristic Rules to Evaluate Commits}
  \label{tab:heuristic}
  \resizebox{\linewidth}{!}{
  \begin{tabular}{clp{12cm}}
    \toprule
    \textbf{Category}&\textbf{Rules}&\textbf{Description}\\
    \midrule
    \multirow{9}{*}{High-impact Rules} & \texttt{\small has\_resolved\_review\_comments} & Evaluate if the commit has resolved review comments. True score: 1 (Receive meaningful feedback and improvements); False score: 0.\\
    &\texttt{\small has\_referenced\_line\_changed\_comments} & Evaluate if the commit has review comments where the referenced lines were changed in the merged commit. True score: 1 (Comments addressed real issues that got fixed); False score: 0.\\
    &\texttt{\small exclude\_merge\_commit} & Evaluate if this is NOT a merge commit (merge commits should have low scores). Commits with "Merge branch" or >2 parents score: 0. Otherwise score: 1.\\
    &\texttt{\small exclude\_base\_merge\_commit} & Evaluate if this is NOT a base/merge commit (score 0 for base and merged commits). Base commits and final merge commits are typically not worthy of review.\\
    \midrule
    \multirow{11}{*}{Medium-impact Rules} & \texttt{\small clear\_commit\_message} & Evaluate the clarity of the commit message based on message length, structure, and content. Give a score from 0 to 1 based on proper message length (10-200 words), structure (multi lines), and meaningful content (include generic words such as ``fix" and ``update").\\
    & \texttt{\small conventional\_commit} & Evaluate if the commit follows conventional commit format. Score 1 to commits start with type(scope): description; else 0.\\
    & \texttt{\small reasonable\_commit\_size} & Evaluate if the commit has a reasonable size. Give a score from 0 to 1 based on the number of changed files and lines (Ideal range: 10-200 lines changed and 1-10 files changed).\\
    & \texttt{\small has\_associated\_review\_comments} & Score 1 if the commit has any associated review comments; else 0. Commits with review comments are likely more significant.\\
    \midrule
    \multirow{9}{*}{Low-impact rules} & \texttt{\small issue\_reference} & Score 1 if the commit message references an issue; else 0.\\
    & \texttt{\small semantic\_commit\_message} & Score of 1 if the commit message is semantic and descriptive; else 0. Checks for action words and descriptive content such as ``add", ``remove" and so on.\\
    & \texttt{\small focused\_file\_changes} & Evaluate if the commit changes are focused on related files. Score from 0 to 1, and higher scores mean that commits that change files in the same directory or with a similar purpose (i.e., fewer changed files).\\
    & \texttt{\small descriptive\_commit\_content} & Evaluate if the commit has more descriptive content beyond just the message. Score from 0 to 1 based on line count, message length, and structured content (marked by lists or sections).\\
  \bottomrule
\end{tabular}
}
\end{table}

\subsubsection{PR Classification}
Since each PR instance contains multiple commits, our goal in this step is to identify high-quality commits within each PR and extract their corresponding high-quality CR-related information. Firstly, we design several heuristic rules to evaluate the commits and give an evaluation score. The details of heuristic rules are shown in Table~\ref{tab:heuristic}. They are categorized into three types: high-impact rules, medium-impact rules, and low-impact rules. A higher impact level signifies greater importance and carries more weight in the scoring (high weight: 3.0, medium weight: 2.0, low weight: 1.0). We assign the highest weight to rules indicating direct, actionable review feedback (e.g., \texttt{has\_resolved\_review\_comments}), as these are the strongest indicators of a meaningful CR process. Rules related to best practices and commit clarity receive medium weight, while secondary indicators like issue references have a low weight. Description presents the scoring details of each heuristic rule. By implementing a weighted sum scoring, we obtain an overall score for each commit. Based on the overall score, we rank the commits in each PR and extract the target commit with the highest score. 

After getting the target commit, high-quality CR task instances can be extracted from the target commit. Specifically, we extract and review comments based on whether referenced lines were actually changed in the merged commit, or the review thread was resolved, outdated, or collapsed. Review comments containing the aforementioned attributes will have their corresponding CRs labeled as high-quality CR task instances.

\subsubsection{Feature Labeling}\label{sec:feature labeling}
In this step, we classify the three attributes: problem domain, difficulty, and review effort. Problem domain refers to the category that the PR's associated issue problem statement belongs to. Referring to SWE-Bench~\citep{jimenez2024}, an issue-solving benchmark, there are nine categories of PR problem domains, as summarized in Table~\ref{tab:problem domain}. Difficulty means how difficult it would be to implement a PR. It can be categorized as low, medium, and high. Review effort means the effort required to review a code change. It is on a scale of 1 to 5. 5 means the most effort. If the repository contains annotations for the above three attributes, we directly utilize the labels from the repository. Otherwise, we employ the LLM-as-a-judge approach~\citep{gu2025}. We leverage the Qwen3-235B-A22B model~\citep{yang2025}, constructing prompts from the Title, Body, Commit Patch to Review, and Head Commit Message fields in Table~\ref{tab:details of fields} as input to classify the three attributes. At last, we get 40,124 CR task instances.

\begin{table}[t]
  \caption{Problem Domain Category}
  \label{tab:problem domain}
  \resizebox{\linewidth}{!}{
  \begin{tabular}{cl}
    \toprule
    \textbf{Problem Domain}&\textbf{Description}\\
    \midrule
    Bug Fixes (BF)& Resolving functional errors, crashes, incorrect outputs\\
    New Feature Additions (NFA) & Adding new functionality or features to the application\\
    Code Refactoring / Architectural Improvement (CA) & Improving code structure, readability, maintainability without changing external behavior\\
    Documentation Update (DU) &Changes related to code comments or external documentation\\
    Test Suite / CI Enhancements (TC)&Improving test coverage, test quality, or continuous integration processes\\
    Performance Optimizations (PO) &Improving application speed, response time, or resource usage efficiency\\
    Security Patches / Vulnerability Fixes (SV) &Fixing code defects that could lead to security issues\\
    Dependency Updates \& Environment Compatibility (DE) &Updating third-party library dependencies or ensuring compatibility across different environments\\
    Code Style, Linting, Formatting Fixes (CLF)& Ensuring code complies with team coding standards and consistency\\
  \bottomrule
\end{tabular}
}
\end{table}

\subsubsection{Manual Selection \& Annotation}
In this step, we manually select high-quality CR task instances that cover nine types of PR problem domains from the 40,124 CR task instances that have passed the filtering process above to serve as the benchmark. Specifically, we invite two of the authors who have more than 5 years of Python development experience to conduct the selection. Firstly, we find that a non-negligible percentage of review comments, while linked to source code lines in the commit patch, are unlikely to
result in code revision in the next round. This kind of review comment is regarded as a noise comment. For example, if a review: (1) that is a simple case (e.g., ``looks good to me'', ``Thanks!'' or ``Nice''); (2) only requests formatting changes with no impact on code logic (e.g., ``fix indentation'', ``add spaces''); (3) requests adding tests (these do not change the code under review); (4) asks for clarification or explanation without suggesting changes (e.g., ``please explain'', ``what does this do?''); (5) refer previous comments that cannot be identified (e.g., ``same as before'', ``see above''); (6) only requests adding comments or documentation (e.g., ``add Javadoc'', ``document this method''); (7) that is difficult to identify (e.g., ``At least here it is clear that the equals method of the implementers of TreeNode is important''), it is considered as a noise and irrelevant review comment. Instances that include these comments are not selected in the final benchmark. In addition, we remove instances whose issue problem statement has images, external hyperlinks, references to specific commit SHAs, and references to other pull requests or issues~\citep{jimenez2024}. The instance with fewer than 40 words in the problem statement is also removed. They are not considered in our scenario. Based on the above rules, we have filtered out 7,086 instances. After that, referring to SWE-Bench~\citep{jimenez2024} and Multi-SWE-Bench~\citep{zan2025}, we design a questionnaire\footnote{To be presented.} about instance quality assessment and ask the two developers to complete the questionnaire for each instance. This questionnaire covers the following areas: (1) Problem statement and patch alignment; (2) Review scope and comment coverage; (3) Defects identified in the patch; (4) Difficulty and review effort; (5) Overall patch quality and risk; (6) Dataset suitability; and (7) Confidence. Thereby, we select the 601 high-quality instances from 70 projects by manual annotation.

\subsection{Benchmark Characteristics}\label{sec:benchmark characteristics}
\subsubsection{Comprehensiveness-aware Benchmark} \system{} is the pioneering CR benchmark to introduce the comprehensiveness-aware concept. Table~\ref{tab:benchmark difference} lists the difference of \system{} and other representative benchmarks presented in Section~\ref{sec:CR related work}, where \CIRCLE~means containing relevant information, \Circle~means containing no relevant information, and \LEFTcircle~means only containing partial relevant information. We highlight a significant oversight in prior research: the comprehensiveness of a CR task. Most of the other CR benchmarks lack basic information and repository-level context information. They only have method-level patches and review comments, which significantly deviate from real CR scenarios. Consequently, they make automatic CR approaches miss some context information that contributes to CR. Further, the performance of LLMs in real-world CR remains unknown. Regarding some issue-solving benchmarks, they also lack some PR and CR-related information. \system{} fills this gap by including basic information, repository-level context information, and more completed PR-related and CR-related information. Thereby, they can achieve repository-level CR. It serves as a benchmark that better simulates real-world CR, aiming to more accurately reflect the true performance of CR approaches.

\subsubsection{Strict Filter Process}
We ensure the quality of the selected CR task instances from several aspects. First, we ensure that the selected projects are of high quality because high-quality projects are generally mature and active. They also have excellent maintenance and a complete PR and CR process. We select Python projects from the top 1,000 projects based on the number of stars, with an average of 21k stars. Also, we keep projects with more than 1,500 PRs, making the collection focus on those with rich, mature, and standardized CR practices. Retaining only projects maintained within the past year ensures that the analysis focuses on ``active'' projects whose CR practices reflect contemporary standards, thereby guaranteeing the research's cutting-edge relevance and practical applicability.

In addition, we ensure that we select merged PRs and PRs with at least 1 closing issue reference. On the one hand, merged PRs indicate that the PR has passed the project's quality gates (including CR and automated testing). On the other hand, PRs associated with closing issues can make sure that every PR has a clear, traceable development intent and rich context.

Furthermore, we design complete heuristic rules to select a representative commit for each PR. These rules can ensure the selected commits: (1) directly reflect collaboration and improvements during CR (e.g., \texttt{\small has\_resolved\_review\_comments}); (2) possess clear, structured, and traceable contextual information (e.g., \texttt{\small conventional\_commit}, \texttt{\small issue\_reference}); (3) adhere to best practices for atomicity and focus (e.g., \texttt{\small reasonable\_commit\_size}, \texttt{\small focused\_file\_changes}); (4) exclude noise commits generated by version control operations that do not reflect development intent (e.g., \texttt{\small exclude\_merge\_commit}). All these rules can ensure the selected commits have high quality.

\section{Evaluation Metric Design}\label{sec:evaluation}

\begin{figure}[t]
  \centering
  \includegraphics[width=0.9\linewidth]{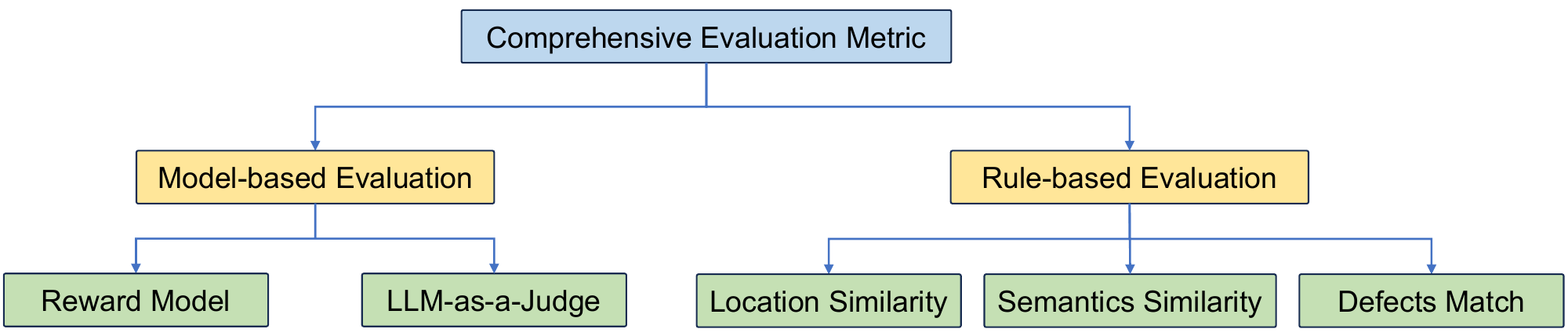}
  \caption{Evaluation Metric Framework}
  \label{fig:evaluation_metric}
\end{figure}

A comprehensive benchmark like \system{} requires an equally comprehensive evaluation framework. Existing evaluation metrics on CR only focus on the method-level patch to review, and they only involve evaluation of semantics similarity and consistency, which can not adapt to comprehensive CR. We propose a comprehensive evaluation metric to evaluate the quality of comprehensive CR. Fig.~\ref{fig:evaluation_metric} shows an overview of the evaluation metric. It consists of two parts, model-based evaluation and rule-based evaluation.

\subsection{Model-base Evaluation}
Model-based evaluation is a black-box evaluation designed to leverage the powerful semantic understanding capabilities of LLMs to score and evaluate the quality of CR. It aims to simulate human judgment of CR's ``semantic quality'' and ``usefulness''. In our paper, we use two kinds of evaluators in model-based evaluation, namely, the reward model and LLM-as-a-judge~\citep{son2024}.

\subsubsection{Reward Model}\label{sec:reward model}
Reward model~\citep{yang2024bayesian} is an AI model that learns human preferences. Its core task is: given an input and one or more model outputs  (Output/Candidate), it outputs a scalar score (Reward Score) that predicts the degree of human preference for that output (i.e., the extent to which it is perceived as high quality). In order to train a reward model to evaluate comprehensive CR, we need to collect the corresponding training set and design the training strategy. 

\textbf{Data Collection}. Firstly, we define the positive CR data and the negative CR data. We collect data from GitHub repositories with over 100 stars. To prevent data leakage, we filter out the repositories used to select CR task instances in Fig.~\ref{fig:pipeline}. We use commits as the collection unit and adopt the \texttt{\small has\_resolved\_review\_comments} and \texttt{\small has\_referenced\_line\_changed\_comments} rules from Table~\ref{tab:heuristic} as labeling criteria. If both rules are True, the commit is classified as positive CR data and labeled as 1. Otherwise, the commit is classified as negative CR data and labeled as 0. At last, we collect 174,661 positive data and 114,458 negative data.

\textbf{Training Strategy}. Our reward model is built upon the Qwen3-8B~\citep{yang2025} pretrained LLM, where we replace the final LM head layer with a Multi-Layer Perceptron consisting of two linear layers and a ReLU activation function. The first linear layer has dimensions of $(\text{hidden\_size}, \text{hidden\_size})$, while the second layer has dimensions of $(\text{hidden\_size}, 1)$. The final output represents a relevance score indicating how well the CR correlates with the given query and patch.

The training loss comprises two components. The first component follows a similar approach to the reward model training in InstructGPT~\citep{ouyang2022training}, employing a contrastive learning-style Bayesian Personalized Ranking (BPR) loss~\citep{rendle2012bpr}. Specifically, for the same query and patch, we sample positive and negative reviews along with their corresponding code context (reviewed file content as default) and diff hunks to achieve this objective. The loss is formulated as:
\begin{equation}
\mathcal{L}_\text{BPR} = -\log \sigma(r_\theta(x, y^+) - r_\theta(x, y^-)).
\end{equation}

Additionally, considering our application scenario of classifying review relevance, we introduce an auxiliary classification objective using binary cross-entropy loss~\citep{wang2025unifiedrewardmodelmultimodal}, that is,
\begin{equation}
\mathcal{L}_\text{BCE} = -\sum_{i=1}^{N} [l_i \log p_i + (1-l_i) \log(1-p_i)],
\end{equation}
where $p_i = \sigma(r_\theta(x_i, y_i))$. Specifically, $r_\theta$ denotes the reward model parameterized by $\theta$, $x$ represents the input consisting of query and patch information, $y^+$ and $y^-$ denote positive and negative reviews respectively, $\sigma$ is the sigmoid function, $l_i \in \{0, 1\}$ is the binary relevance label for the $i$-th sample, and $p_i$ is the predicted probability of relevance. The total training loss is the weighted combination of these two components:
\begin{equation}
\mathcal{L}_\text{total} = \mathcal{L}_\text{BPR} + \lambda \mathcal{L}_\text{BCE},
\end{equation}
where $\lambda$ controls the relative importance of $\mathcal{L}_\text{BCE}$. We set it to $1$ by default.

\textbf{Implementation Details}. For efficient training, we employ Low-Rank Adaptation (LoRA)~\citep{hu2022lora} as our parameter-efficient fine-tuning method with a rank of 32 and an alpha value of 16, which significantly reduces the number of trainable parameters while maintaining model performance. The model is trained for 1 epoch with a learning rate of $5e-5$. To accommodate the comprehensive CR context, we set the maximum sequence length to 24,576 tokens, enabling the model to process lengthy code patches and associated reviews effectively.

We utilize Flash Attention 2~\citep{dao2023flashattention2} for memory-efficient attention computation, with mixed precision training with bfloat16. The training process incorporates 100 warm-up steps for learning rate scheduling to ensure stable convergence. The distributed training is conducted using DeepSpeed ZeRO~\citep{rajbhandari2020zero} Stage 2. All experiments are conducted on 32 NVIDIA A100 GPUs with a total effective batch size of 64.

\subsubsection{LLM-as-a-Judge}
\begin{figure}[t]
  \centering
  \includegraphics[width=0.9\linewidth]{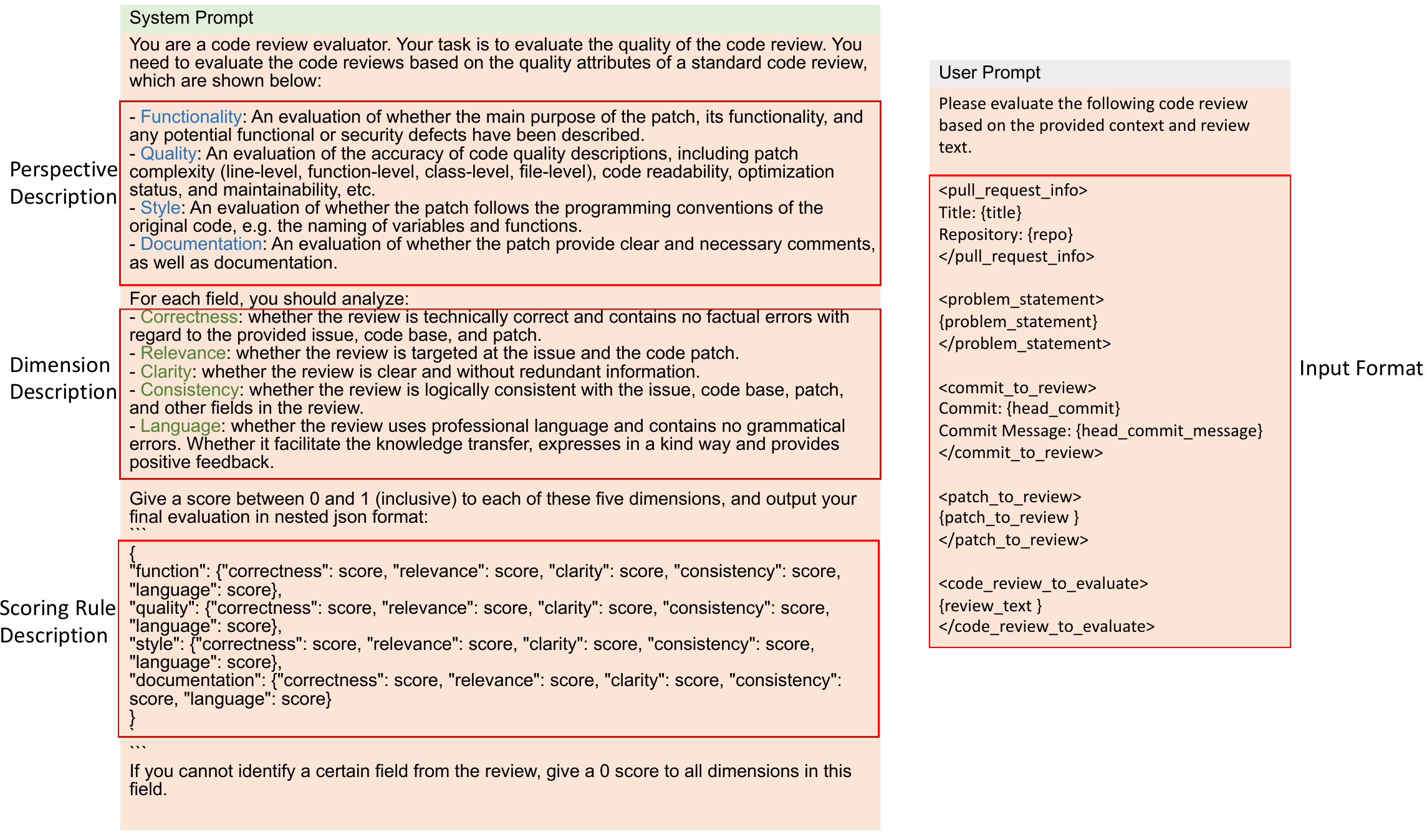}
  \caption{LLM-as-a-Judge Prompt}
  \label{fig:llm_prompt}
\end{figure}

To evaluate the CR in a general perspective, we also adopt LLM-as-a-judge to evaluate the CR quality. Specifically, we design a prompt that consists of CR information and evaluation criteria to request an LLM to provide a score for the CR quality. Referring to some research about CR usefulness investigation~\citep{Yang2023,turzo2024}, we request LLM to provide a score of CR based on four perspectives, that is, Functionality, Quality, Style, and Documentation. For each perspective, we request LLM to analyze five dimensions of Correctness, Relevance, Clarity, Consistency and Language. Each dimension will be provided a score by LLM and then an average overall score will be calculated at last. The detail prompt is shown in Fig.~\ref{fig:llm_prompt}. We use OpenAI o3-2025-04-16~\citep{o3} model to achieve LLM-as-a-judge. Thereby, we can obtain a score provided by LLM.

By combining scores based on the reward model and LLM-as-a-judge, respectively, we get an average score with model-based evaluation.

\subsection{Rule-base Evaluation}
To achieve a comprehensive evaluation metric, we also adopt a rule-based evaluation. Rule-based evaluation is a white-box evaluation, which aims to determine the ``formal correctness'' and ``superficial similarity'' of CR. We automatically extract structured defect information from the CR reports generated by the LLMs, including file paths, line numbers, and review comments. Then we conduct some heuristic rules which consists of location similarity, semantics similarity and defects match to implement rule-based evaluation.
\subsubsection{Location Similarity}Location similarity aims to evaluate the matching between the predicted CR location and the CR ground truth. It includes file path matching, line number proximity, and diff hunk match level. Specifically, if the file path where the predicted CR locates is exact match with the ground truth, we give a score of 1. Otherwise, we give a score of 0. Regarding the line number accuracy, if it is an exact match with the ground truth, we give a score of 1. If the line difference $\mathrm{diff}_\text{line}$ is more than 5 lines, we give a score of 0.1. Otherwise, we conduct a decay function to calculate the line number accuracy $\mathrm{Acc_{LN}}$, which is shown as follows.
\begin{equation}
\label{eq:line number}
\mathrm{Acc_{LN}}=\left\{
\begin{aligned}
1 \quad\quad\quad & , & \mathrm{diff}_\text{line} = 0, \\
1-\frac{\mathrm{diff}_\text{line}}{10} & , & \mathrm{diff}_\text{line} \leq 5,\\
0.1 \quad\quad\quad & , & \mathrm{diff}_\text{line} > 5.\\
\end{aligned}
\right.
\end{equation}
In addition, to avoid unduly penalizing essentially correct predictions due to minor line number offsets, we introduce a more robust metric: diff hunk similarity. This metric evaluates whether the LLM's predicted CR location and the ground truth location reside within the same code change block. Specifically, We analyze whether the predicted CR's line number resides within the same diff hunk as the ground truth. If it does, we assign a score of 1. If the predicted CR is in a different diff hunk, we assign a score of 0. If the predicted CR's line number is not present in any diff hunk in the patch to review, we collect the five lines above and below the CR line and calculate the overlap ratio between this range and the ground truth diff hunk's line number range as the diff hunk similarity score. Thereby, combining with the file path matching and line number accuracy, diff hunk similarity can contribute to a more comprehensive and fairer location evaluation. We assign weights of 70\%, 15\%, and 15\% to file path matching, line number accuracy, and diff hunk similarity, respectively, ultimately yielding a composite location similarity score.

\subsubsection{Semantics Similarity}
In order to evaluate the semantics performance of review comments, we use BLEU~\citep{Papineni02}, a popular evaluation metric that is implemented in neural machine translation and conversation
systems. BLEU is now widely used in code-related tasks, such as code comment generation~\citep{Guo2023}, document generation~\citep{Hu2022}, review comment generation~\citep{Li2022} and so on. Its core idea is to compare the n-gram overlap between the text generated by the model and the ground truth. The greater the overlap, the higher the BLEU score. The specific calculation process of BLEU is given as follows:
\begin{equation}
  \mathrm{BLEU} = \mathrm{BP} \cdot \exp(\sum_{n=1}^{m}\omega_{n} \log p_{n}),
\end{equation}
where $\omega_{n}$ is the weight of $n$-gram and $p_{n}$ is the precision of $n$-gram. Usually, the maximum value of $n$ is 4, which is represented by BLEU-4. $\mathrm{BP}$ is the brevity penalty factor for generated review comment length, which is shown as follows: 
\begin{equation}
\mathrm{BP}=\left\{
\begin{aligned}
1 \quad\quad & , & l_{c}> l_{r}, \\
\exp(1-\frac{l_{r}}{l_{c}}) & , & l_{c}\leq l_{r}.
\end{aligned}
\right.
\end{equation}
where $l_{c}$ is the length of generated review comment and $l_{r}$ represents the reference review comment. In this paper, we employ BLEU-4 as the semantics similarity. 

\subsubsection{Defects Match}
Since a CR task may involve reviewing multiple review comments and locations within a commit, we design a defect match rule to evaluate performance under multi-location review scenarios. Specifically, for each predicted review comment and location (i.e., predicted defect), we calculate the average of the location similarity and semantics similarity scores between it and each ground-truth review comment and location, namely sub-defect-match scores. We select the highest sub-defect-match score among them. If this score exceeds the set threshold, we consider the predicted review comment and location to be correctly matched to the defect information. Based on the number of correctly matched defects, we can calculate precision, recall, and F1 scores.
\begin{equation}
\mathrm{Precision} = \frac{N_\text{correctly\_matched\_defects}}{N_\text{predicted\_defects}},
\end{equation}
\begin{equation}
\mathrm{Recall} = \frac{N_\text{correctly\_matched\_defects}}{N_\text{ground\_truth\_defects}},
\end{equation}
\begin{equation}
F1 = \frac{2 * \mathrm{Precision} * \mathrm{Recall}}{\mathrm{Precision}+\mathrm{Recall}}.
\end{equation}

The total score of defect match is calculated by averaging the F1 score and the average of sub-defect-match scores. After getting the total defect score, we combine and average it with location similarity and semantics similarity score to get a rule-based score. At last we get an overall evaluation score based on the average of the model-based score and the rule-based score.

\section{Experiment Design}
In this section, we first provide the detailed distribution of the benchmark, introduce the models, context acquisition strategy, and the prompt used in the experiments, and provide a detailed description of the experimental setups. 
After that, we answer the following four research questions that the experiments aim to address:
 \begin{enumerate}[leftmargin=*,start=1,label={\bfseries RQ\arabic*:}]
\item What is the LLMs' performance of comprehensive CR in \system{} benchmark?
\item What is the LLM's performance of comprehensive CR in different PR types?
\item How does different contextual information contribute to CR performance?
\item What is the performance of the reward model?
\end{enumerate}

\begin{figure}[t]
  \centering
  \begin{subfigure}[t]{0.45\textwidth}
    \centering
    \includegraphics[width=\linewidth]{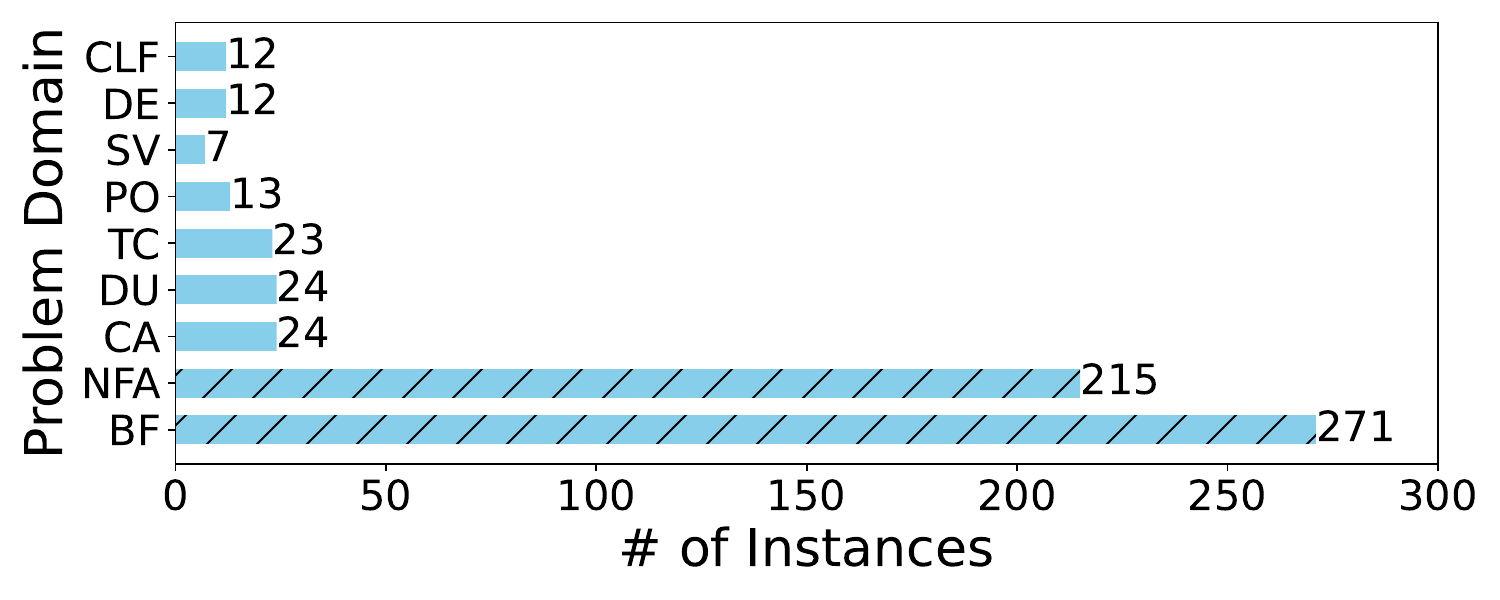}
    \caption{Problem Domain Distribution}
    \label{fig:Domain Distribution}
  \end{subfigure}
  \hfill
  \begin{subfigure}[t]{0.52\textwidth}
    \centering
    \includegraphics[width=\linewidth]{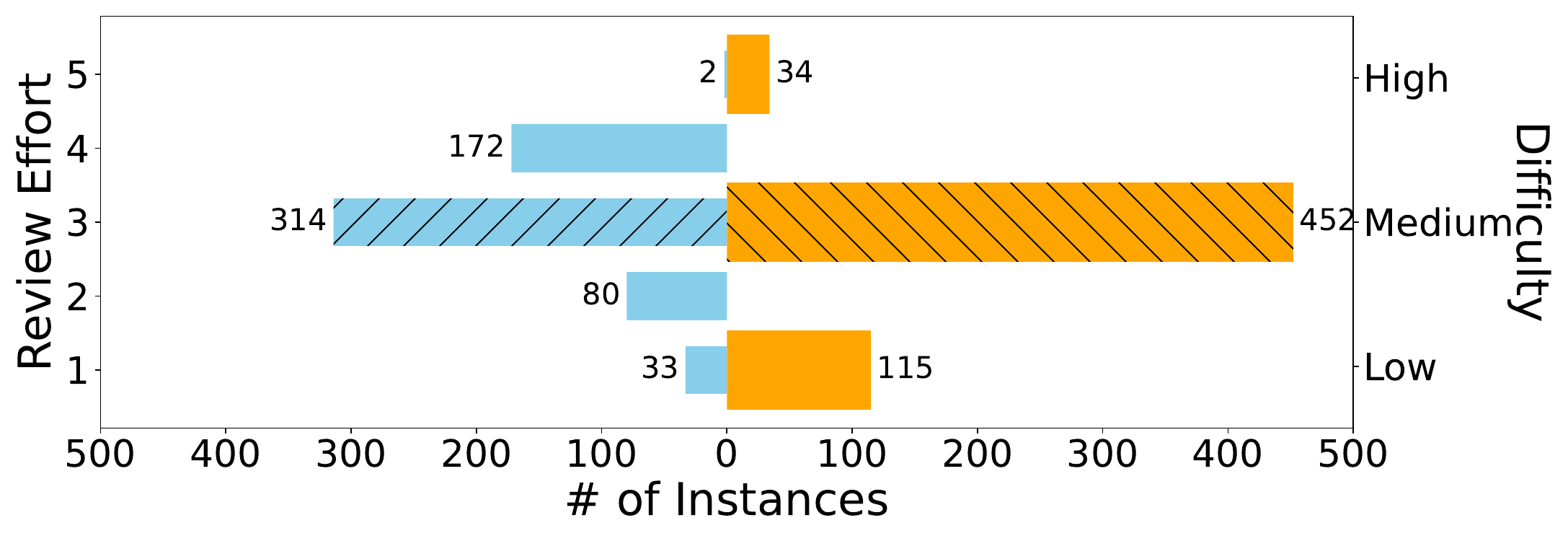}
    \caption{Review Effort \& Difficulty Distribution}
    \label{fig:effort and difficulty distribution}
  \end{subfigure}

  \caption{Benchmark Distribution}
  \label{fig:twosubfigs}
\end{figure}

\subsection{Benchmark Distribution}

After implementing the pipeline in Section~\ref{sec:benchmark construction}, we construct a benchmark with 601 Python CR task instances, which were created from 2016-11-06 to 2025-07-07. The problem domain distribution is shown in Fig.~\ref{fig:Domain Distribution}. It can be found that bug fixes constitute the problem domain with the highest volume of high-quality CRs. This indicates robust software maintenance practices and a thriving CR ecosystem for this category of issues. The review effort and difficulty distribution are shown in Fig.~\ref{fig:effort and difficulty distribution}. Most instances are in a medium difficulty of PR task implementation. The most common review effort is 3, which also indicates that most CRs are assigned to tasks of moderate difficulty.

\begin{wrapfigure}{r}{0.35\textwidth}
\centering
  \includegraphics[width=\linewidth]{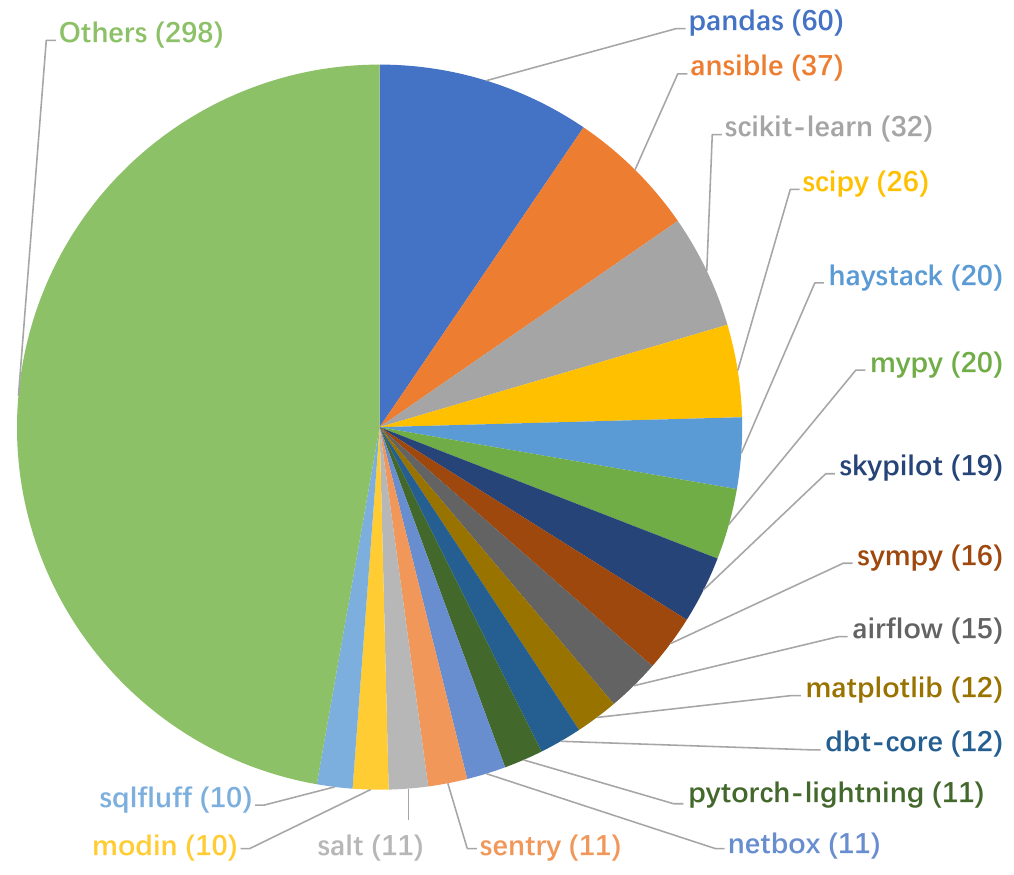}
  \caption{Project Distribution}
  \label{fig:project_distribution}
\end{wrapfigure}

The 601 instances involve 70 Python projects. Fig.~\ref{fig:project_distribution} illustrates the project distribution of the benchmark. We present the top 17 projects containing the highest number of instances. These projects account for 55.41\% of the total instances. Most of these projects are well-known within the Python community (e.g., pandas, scikit-learn et al.). This indicates that these projects are well-maintained with high-quality CRs.

\subsection{Studied LLMs}

We consider mainstream LLMs—encompassing both open-source and proprietary models—that have been widely adopted in recent software-engineering-related studies~\citep{Fan2025, Batole2025, Wang2024, Yin2024}. For open-source LLMs, we select DeepSeek-v3.1~\citep{guo2025deepseek},
Kimi-K2-0905-preview~\citep{kimiteam2025kimik2openagentic}
Qwen3-235B-A22B~\citep{yang2025}. For closed-source LLMs, we choose the commonly used commercial models: Claude-Sonnet-4-20250514~\citep{claude4}, Gemini 2.5 Pro~\citep{comanici2025gemini}, GPT-4o~\citep{hurst2024gpt}, GPT-5~\citep{GPT-5}. They have excellent performance in code-related tasks in previous benchmarks.

\subsection{Studied Context Acquisition Strategies}\label{sec:context strategy}
As a repository-level CR benchmark, \system{} can provide repository-level contextual information as part of the LLM input to help the LLM achieve comprehensive CR. Due to the large size of modern software repositories, it
is not feasible to feed the entire repository to the model. It is necessary to retrieve the most relevant information. Referring to SWE-Bench~\citep{jimenez2024}, we adopt two types of context acquisition strategies that might be beneficial for comprehensive CR: Retrieval-based Acquisition and Oracle-based context acquisition. In RQ3, we will validate the performance of these two strategies.

\textbf{Retrieval-based Context Acquisition}.
Since the problem description and the code implementing the functionality may share overlapping vocabulary, we employ BM25~{Stephen2004} to retrieve code files similar to the problem statement within the project. BM25 is a ranking function commonly used in information retrieval to measure the relevance between a document and a query. As a kind of sparse retrieval method, BM25 offers fast computation speed and low resource consumption, making it highly suitable for file retrieval scenarios in large-scale software projects. In RQ3, we will separately analyze the impact of using the top-1, top-3 and top-5 relevant code files retrieved via BM25 as context on the effectiveness of LLM-based CR.

\textbf{Oracle-based Context Acquisition}.
As a baseline, we also consider the oracle-based retrieval. Specifically, we retrieve the combination of changed files in $\mathrm{diff}(base\ commit, commit\ to\ review) \cup \mathrm{diff}(base\ commit, merged\ commit)$. The reason lies in the fact that there are some other commits related to the target commit in the PR when extracting the target commit in PR as the CR target. These related commits may serve as supplementary modifications to the target commit. Therefore, we choose the change files from commits in the same PR as the largest context space.

\subsection{Experimental Setting}
\begin{wrapfigure}{r}{0.52\textwidth}
\centering
  \includegraphics[width=\linewidth]{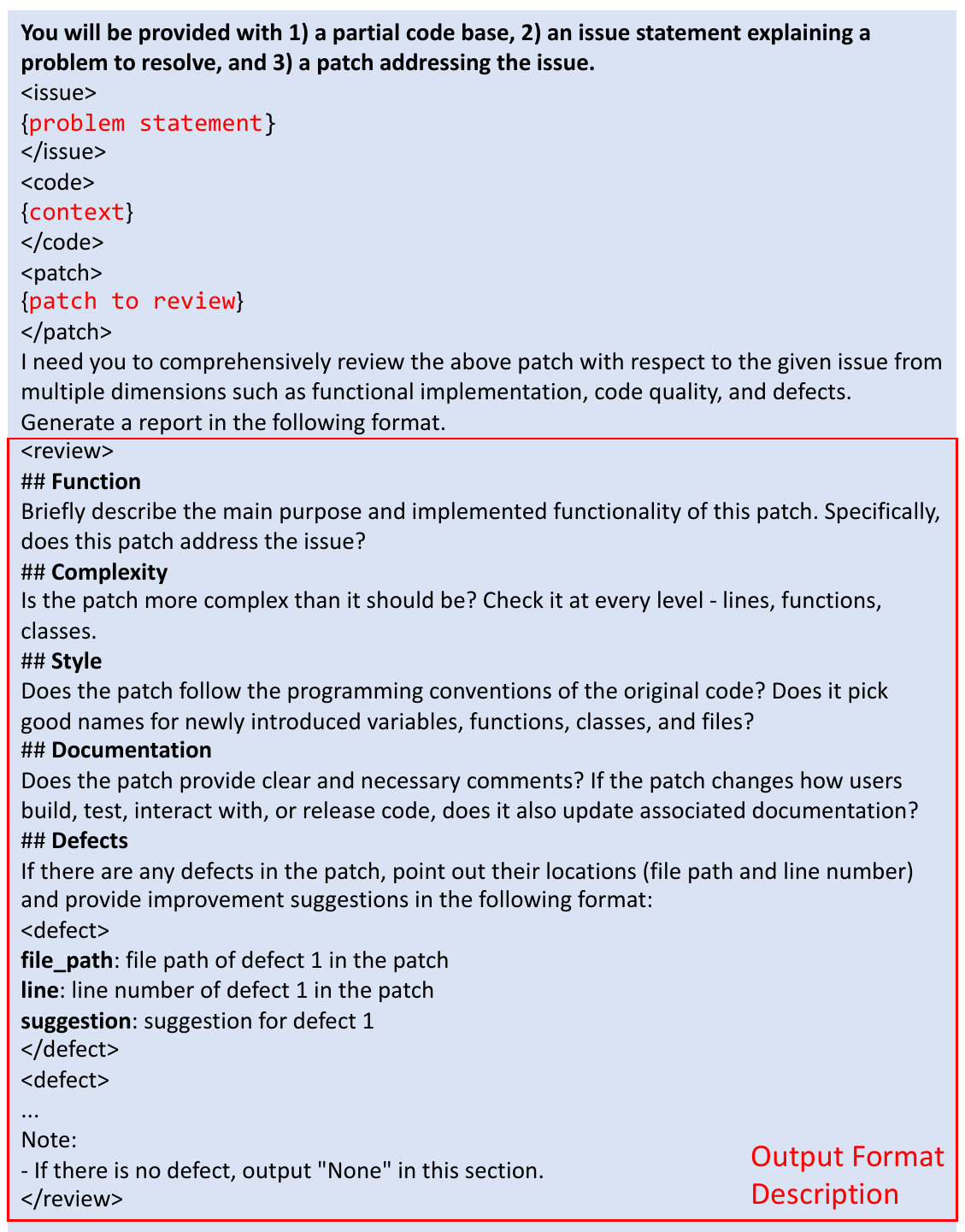}
  \caption{Prompt Template}
  \label{fig:prompt_template}
\end{wrapfigure}

We use the evaluation framework proposed in Section~\ref{sec:evaluation} to assess the CR generated by LLMs. We set the generation temperature of LLMs to 0.6 (except GPT-5 for 1), top-p to 0.95. The context window is set based on the maximum context window length specified in each LLM's system card. Note that, when employing different context retrieval strategies, once the token length exceeds the context window, we remove the retrieved files outside the context window. When the input remains longer than the context window length after removing all retrieved files, we consider the LLM unable to provide a correct CR for that instance and assign an evaluation score of 0. To mitigate issues stemming from the randomness of model generation, the experimental results presented in this paper are obtained by conducting three repeated experiments and averaging the results. The prompt template is shown in Fig.~\ref{fig:prompt_template}, which consists of the \texttt{\small issue} tag including the problem statement, the \texttt{\small code} tag including the context, and the \texttt{\small patch} tag including the patch to review. The generated CR format is also described in the \texttt{\small review} tag, which involves the functional implementation, code quality, and defect information.

\subsection{RQ1: What is the LLMs' performance of comprehensive CR in \system{} benchmark?}

We conduct experiments under the oracle-based context as the baseline setting to explore the performance of mainstream LLMs in comprehensive CR. Table~\ref{tab:llm performance} shows the results, which reveal distinct strengths and weaknesses across models. This suggests that Gemini is currently the most well-rounded performer in generating contextually grounded and technically sound CR feedback.

On the model-based evaluation dimension, GPT-5 attains the best score (64.80), followed closely by Gemini 2.5 Pro (63.65). This demonstrates that GPT-5 generates review comments with the highest perceived quality in terms of semantic quality and usefulness, even if its overall integration across dimensions is not optimal.

In contrast, Claude-Sonnet-4-20250514 excels in the rule-based evaluation (33.31) and significantly outperforms models like GPT-4o (8.10) and GPT-5 (18.30), indicating its strong adherence to location similarity, semantics similarity, and defect match. This highlights Claude’s strength in producing formally correct and precise feedback, albeit sometimes at the cost of broader contextual integration.

Notably, several models exhibit significant imbalances. For instance, while GPT-5 performs best in model-based scoring, its rule-based performance lags, resulting in a lower comprehensive score (41.96) than Gemini. Similarly, GPT-4o shows moderate model-based performance but severely underperforms in rule-based checks, suggesting potential issues with factual grounding or hallucination in CR scenarios. These findings underscore that no single LLM dominates across all dimensions of comprehensive code review. Instead, performance varies significantly depending on the evaluation axis, emphasizing the necessity of our multi-faceted assessment framework. The results also highlight the importance of integrating both model-based and rule-based metrics to avoid overestimating the practical utility of generated CRs.

\begin{table}
  \caption{LLM Performance Score (\%) under \system{} Banchmark}
  \label{tab:llm performance}
  \footnotesize
  \renewcommand{\arraystretch}{0.85}
  \begin{tabular}{lccc}
    \toprule
    \textbf{LLM} & \textbf{Model-based Score} & \textbf{Rule-based Score} & \textbf{Comprehensive Score}\\
    \midrule
    DeepSeek-v3.1 & 58.69 & 28.34 & 42.51 \\
    Kimi-K2-0905-preview & 62.11 & 20.81 & 46.77\\
    Qwen3-235B-A22B & 58.10 & 24.30 & 40.45\\
    Claude-Sonnet-4-20250514 & 60.67 & \textbf{\underline{33.31}} & 47.46\\
        Gemini 2.5 Pro & 63.65 & 29.47 & \textbf{\underline{52.37}}\\
    GPT-4o & 54.57 & 8.10 & 35.47\\
    GPT-5 & \textbf{\underline{64.80}} & 18.30 & 41.96\\
  \bottomrule
\end{tabular}
\end{table}

\begin{center}
\fcolorbox{black}{gray!10}{\parbox{\linewidth}{\textbf{Answer to RQ1}: Gemini 2.5 Pro achieves the highest comprehensive CR performance on \system{}, outperforming other LLMs by balancing strong model-based and rule-based scores, while GPT-5 and Claude-Sonnet excel in semantic quality and formal correctness, respectively, highlighting the necessity of multi-dimensional evaluation for realistic CR assessment.}}
\end{center}

\subsection{RQ2: What is the LLM's performance of comprehensive CR in different PR types?}

In Section~\ref{sec:feature labeling}, we label the CR task instances into nine problem domains based on the PR types. In this RQ, we conduct a fine-grained analysis of their comprehensive CR capabilities within each domain, with results presented in Table~\ref{tab:problem domain score}. It can be found that Gemini 2.5 Pro achieves the highest comprehensive score in all nine problem domains. It outperforms the second-best model in each domain by margins ranging from 5.93\% to 28.17\%, with particularly strong performance in DU (58.86), TC (55.83), and DE (53.35).

This uniform dominance suggests that Gemini 2.5 Pro not only generalizes well across different types of code changes but also effectively leverages domain-specific context—such as test requirements, security constraints, or formatting rules—to generate more accurate and actionable CR feedback. The results highlight its robustness and adaptability in handling the heterogeneous nature of modern CR practices.

\begin{table}
  \caption{LLM's Comprehensive Score (\%) in Different Problem Domains}
  \label{tab:problem domain score}
  \footnotesize
  \renewcommand{\arraystretch}{0.85}
  \begin{tabular}{lccccccccc}
    \toprule
    \textbf{LLM} & \textbf{BF} & \textbf{NFA} & \textbf{CA} & \textbf{DU} & \textbf{TC} & \textbf{PO} & \textbf{SV} & \textbf{DE} & \textbf{CLF}\\
    \midrule
    DeepSeek-v3.1 & 43.24 & 41.75 & 38.73 & 47.20 & 47.98 & 42.62 & 38.45 & 47.44 & 24.54\\
    Kimi-K2-0905-preview & 46.34 & 47.09 & 46.68 & 47.43 & 51.20 & 49.24 & 43.36 & 47.19 & 39.97\\
    Qwen3-235B-A22B & 41.39 & 40.14 & 30.47 & 45.31 & 46.91 & 37.11 & 35.23 & 46.91 & 22.92\\
    Claude-Sonnet-4-20250514 & 49.26 & 46.42 & 47.04 & 47.27 & 46.66 & 44.48 & 39.21 & 49.44 & 33.94\\
    Gemini 2.5 Pro & \textbf{\underline{52.14}} & \textbf{\underline{51.93}} & \textbf{\underline{49.83}} & \textbf{\underline{58.86}} & \textbf{\underline{55.83}} & \textbf{\underline{51.62}} & \textbf{\underline{52.11}} & \textbf{\underline{53.35}} & \textbf{\underline{51.23}}\\
    GPT-4o & 36.44 & 35.23 & 32.09 & 37.66 & 36.00 & 37.94 & 31.93 & 35.33 & 18.49\\
    GPT-5 & 43.67 & 40.68 & 36.73 & 44.88 & 44.25 & 43.04 & 36.43 & 44.16 & 26.17\\
  \bottomrule
\end{tabular}
\end{table}

\begin{center}
\fcolorbox{black}{gray!10}{\parbox{\linewidth}{\textbf{Answer to RQ2}: Gemini 2.5 Pro consistently outperforms all other LLMs in comprehensive CR across all nine problem domains, demonstrating its superior generalization and context utilization in diverse CR tasks.}}
\end{center}

\subsection{RQ3: How does different contextual information contribute to CR performance?}

In Section~\ref{sec:context strategy}, we propose two types of context acquisition strategies that may be helpful for LLMs to achieve comprehensive CR. In this RQ, we evaluate the effectiveness of these strategies under varying context availability. Particularly, regarding retrieval-based context acquisition, we adopt BM25 to retrieve top-1, top-3 and top-5 relevant files as the context, respectively. 

Results, as shown in Table~\ref{tab:context score}, demonstrate that Gemini 2.5 Pro achieves the highest comprehensive score across all context configurations, including both oracle-based and BM25-based settings. Notably, its performance under the BM25 top-1 context (52.24) is nearly on par with that under the oracle-based setting (52.37), with only a marginal gap of 0.13. This indicates that Gemini can achieve near-optimal review quality by leveraging just the single most relevant retrieved file, highlighting its strong capability in contextual relevance filtering and efficient information utilization.

Furthermore, Gemini maintains stable performance across different retrieval depths (top-1: 52.24, top-3: 51.45, top-5: 51.63), suggesting robustness to context size and resilience to potential noise in larger retrieval sets. This stability contrasts with other models—such as GPT-5 and GPT-4o—whose scores fluctuate more significantly with context size, indicating higher sensitivity to irrelevant or redundant information. These findings also Gemini 2.5 Pro is the most context-efficient and robust model for practical, scalable CR systems where oracle-level context is unavailable.

\begin{table}
  \caption{LLM's Comprehensive Score (\%) in Different Context}
  \label{tab:context score}
  \footnotesize
  \renewcommand{\arraystretch}{0.85}
  \begin{tabular}{lccccccccc}
    \toprule
    \textbf{LLM} & {\tabincell{c}{\textbf{Oracle-based}\\\textbf{Context}}} & {\tabincell{c}{\textbf{BM25-based}\\\textbf{ Context Top-1}}} & {\tabincell{c}{\textbf{BM25-based}\\\textbf{ Context Top-3}}} & {\tabincell{c}{\textbf{BM25-based}\\\textbf{ Context Top-5}}}\\
    \midrule
    DeepSeek-v3.1 & 42.51 & 48.13 & 43.07 & 36.42\\
    Kimi-K2-0905-preview & 46.77 & 48.15 & 47.78 & 45.43\\
    Qwen3-235B-A22B & 40.45 & 47.29 & 40.71 & 33.55\\
    Claude-Sonnet-4-20250514 & 47.46 & 51.02 & 48.43 & 41.91\\
    Gemini 2.5 Pro & \textbf{\underline{52.37}} & \textbf{\underline{52.24}} & \textbf{\underline{51.45}} & \textbf{\underline{51.63}} \\
    GPT-4o & 35.47 & 39.34 & 36.03 & 31.13\\
    GPT-5 & 41.96 & 44.75 & 49.87 & 40.17\\
  \bottomrule
\end{tabular}
\end{table}

\begin{center}
\fcolorbox{black}{gray!10}{\parbox{\linewidth}{\textbf{Answer to RQ3}: Gemini 2.5 Pro achieves near-oracle performance in comprehensive code review using only the top-1 retrieved context, demonstrating superior context efficiency and robustness across retrieval-based and oracle-based acquisition strategies.}}
\end{center}

\subsection{RQ4: What is the performance of the reward model?}

In Section~\ref{sec:reward model}, we incorporate a reward model as a key component of our model-based evaluation framework. To assess its reliability and validity in distinguishing high-quality from low-quality code reviews, we conduct a comparison study to evaluate the classification performance of the reward model. The input to the reward model consists of two elements: the query and the patch to review. The query itself is composed of the context (i.e., the full content of the reviewed file, used as default context), review context, and the review location information. We split the dataset described in Section~\ref{sec:reward model} into a 90\% training set and a 10\% test set to evaluate the model’s classification effectiveness. A CR is classified as positive if the reward model outputs a score greater than 0.5; otherwise, it is classified as negative. As shown in Table~\ref{tab:reward model evaluation}. the reward model achieves an accuracy of 75.03\% and an F1 score of 80.64\%, demonstrating its strong discriminative capability. Notably, even when the context information is removed, the model maintains competitive performance, with accuracy and F1 scores of 74.30\% and 79.57\%, respectively—suggesting that the review context, location and patch already provide substantial signal for review quality assessment. For comparison, we evaluate two SOTA LLMs—Kimi-K2-0711-preview and Gemini 2.5 Pro—on the same binary classification task. The prompt provided to each LLM mirrors the input format used by the reward model, and the LLM’s output is directly mapped to a binary label based on whether it expresses approval or disapproval of the review. The results show significantly lower performance: Kimi achieves 43.68\% accuracy and 49.58\% F1, while Gemini reaches 51.30\% accuracy and 61.36\% F1. These findings indicate that the fine-tuned reward model substantially outperforms general-purpose LLMs in this specialized classification task, validating its suitability as a reliable evaluator within our comprehensive CR assessment framework. In our final evaluation pipeline, the reward model is applied with the reviewed file context to ensure maximal fidelity to real review scenarios.

\begin{table}
  \caption{Reward Model Performance}
  \label{tab:reward model evaluation}
  \footnotesize
  \renewcommand{\arraystretch}{0.85}
  \begin{tabular}{lccccccccc}
    \toprule
    \textbf{Model Type} & \textbf{Accuracy (\%)} & \textbf{Precision  (\%)} & \textbf{Recall (\%)} & \textbf{F1 (\%)}\\
    \midrule
    Reward Model w Reviewed File & 75.03 & 79.40 & 81.92 & 80.64\\
    Reward Model w/o Context & 74.30& 80.02 & 79.30 & 79.57\\
    Kimi-K2-0711-preview w Reviewed File & 43.68& 59.13&42.69&49.58\\
    Gemini 2.5 Pro w Reviewed File &51.30 & 63.22&59.60&61.36\\
  \bottomrule
\end{tabular}
\end{table}

\begin{center}
\fcolorbox{black}{gray!10}{\parbox{\linewidth}{\textbf{Answer to RQ4}: The reward model we construct demonstrates objective performance in evaluating the quality of CRs, achieving an F1 score exceeding 80\% and an accuracy exceeding 75\%.}}
\end{center}

\section{Threats to Validity}

\textbf{Threats to Internal Validity.}
For threats to internal validity, the first concern is the selection of prompts. The performance of LLMs is sensitive to prompt design. Using an equivalent but differently phrased prompt may result in significant performance differences. To mitigate this threat, we ultimately chose the one that yielded the best results after trying several formats. However, since we did not cover all available prompt formats, the one we selected may not be the optimal one. Secondly, since the setting of LLM hyperparameters (e.g., temperature), the outputs of LLM may exhibit randomness. To mitigate this issue, we conduct three repeated experiments
and average the results. 

\textbf{Threats to External Validity.}
The first threat to external validity is the programming language generalization. \system{} is a benchmark based on Python. So it can not be used to evaluate comprehensive CR on other programming languages such as Java or C++. In the future, we plan to address this limitation by continuously expanding \system{} to cover as many programming languages as possible. The second external validity is the type of generalization. \system{} consists of instances including nine types of problem domains. However, there may exist some CRs involving some other types. We will try to expand the number of types in the future. Another threat is the model generalization. Due to limited resources, our experiments are not able to cover all the available LLMs, thus not fully reflecting the performance of all LLMs in actual development scenarios, which may slightly affect the representativeness of our experiments.

\section{Conclusion}
CR is an important component for building stable, maintainable, and high-quality software products. Many CR benchmarks have been proposed to evaluate different kinds of automatic CR approaches. But there is a lack of comprehensiveness in existing benchmarks and approaches, which are detached from real CR scenarios. In this paper, we introduce \system{}, a comprehensiveness-aware CR benchmark to fill this gap. The design of the benchmark can address the limitations of real-world CR. Correspondingly, we design an evaluation metric to achieve comprehensive CR evaluation. We conduct an  
empirical study to explore SOTA LLMs’ performance in the real-world CR scenario. This more realistic CR benchmark encourages creative automatic CR solutions that can have immediate applicability in open-source software development. We
believe that \system{} will offer
valuable evaluation for the future development of LLM-based automatic CR approaches.

\bibliographystyle{colm2024_conference}
\bibliography{reference}

\appendix
\clearpage

\end{document}